\documentclass[%
 reprint,
superscriptaddress,
showpacs,preprintnumbers,
 amsmath,amssymb,
 aps,
prb,
]{revtex4-1}

\usepackage{comment,color}
\usepackage{braket}

\usepackage{graphicx}% Include figure files
\usepackage{dcolumn}% Align table columns on decimal point
\usepackage{bm}% bold math
\begin{document}

%\preprint{APS/123-QED}

\title{Universal phase transition and band structures for spinless nodal-line and Weyl semimetals %between Topological nodal-line semimetal and Weyl semimetal in spinless systems
}% Force line breaks with \\
%\thanks{A footnote to the article title}%

\author{Ryo Okugawa}
\affiliation{%
 Department of Physics, Tokyo Institute of Technology, 2-12-1 Ookayama, Meguro-ku, Tokyo 152-8551, Japan
}%
\author{Shuichi Murakami}%
\affiliation{%
 Department of Physics, Tokyo Institute of Technology, 2-12-1 Ookayama, Meguro-ku, Tokyo 152-8551, Japan
}%
\affiliation{%
 TIES, Tokyo Institute of Technology, 2-12-1 Ookayama, Meguro-ku, Tokyo 152-8551, Japan
}%

\date{\today}% It is always \today, today,
             %  but any date may be explicitly specified

\begin{abstract}
We study a general phase transition between spinless topological nodal-line semimetal and Weyl semimetal phases.
We classify topological nodal  lines into two types based on their positions and shapes, and their phase
transitions depends on their types. 
We show that a topological nodal-line semimetal becomes the Weyl semimetal by breaking time-reversal symmetry
when the nodal lines enclose time-reversal invariant momenta (type-A nodal lines). 
We also discuss an effect of crystallographic symmetries determining the band structure of the topological nodal-line semimetals.
Thanks to protection by the symmetries, the topological nodal-line semimetals can transition into spinless Weyl semimetals 
or maintain the nodal lines in many crystals after inversion symmetry is broken.
\end{abstract}

%Surface states, band structures, electron density of states,  Quantum phase transitions,  
% PACS, the Physics and Astronomy
                             % Classification Scheme.

\maketitle

\section{INTRODUCTION}
Many topological semimetals are realized by strong spin-orbit interactions leading to the gap closing.
One example of the topological semimetals is a Weyl semimetal (WSM) \cite{Murakami07, Wan11}.
The WSMs have three-dimensional nondegenerate Dirac cones.
The gapless points called Weyl nodes are protected by the topology in the momentum space,
and necessarily appear in pairs.
The WSMs also show topological surface states called Fermi arcs \cite{Wan11, Ojanen13, Okugawa14, Haldane14}.
As another example, topological Dirac semimetals have been also investigated \cite{Wang12,Wang13,Yang14,Fang15}.
The topological Dirac semimetals can be obtained in systems with time-reversal (TR) and inversion (I) symmetries,
and have Dirac nodes with fourfold degeneracy. %due to the Kramer's degeneracy.
Moreover, recent works propose new topological semimetals
whose band crossings lie on symmetry points with high-dimensional irreducible representations \cite{Wieder16, Zhu16, Weng16, Bradlyn16}.

On the other hand, a novel topological semimetal appears in spinless systems with TR- and I symmetries. 
It is called a topological nodal line (TNL) semimetal (SM). 
The TNLSMs have twofold degenerate nodal lines on general points in the three-dimensional Brillouin zone.
(If spin degeneracy is considered, the TNLs are fourfold degenerate.)
The line degeneracy is acccidental, and characterized by a quantized Berry phase equal to $\pi$ \cite{Mikitik99, Heikkila15}.
Hence, the nodal lines are protected topologically.
However, existence of characteristic surface states called drumhead surface states is not necessarily guaranteed\cite{Fang15, Hirayama17}.
The TNLSMs have been predicted theoretically in various materials with negligible spin-orbit interaction.
The candidates are carbon allotropes \cite{Chen15, Weng15b, Wang16}, Cu$_3$(Pd, Zn)N \cite{Kim15, Yu15}, Ca$_3$P$_2$ \cite{Xie15, Chan16}, LaN \cite{Zeng15}, 
compressed black phosphorus \cite{Zhao16}, alkaline-earth metals \cite{Hirayama17, Li16}, BaSn$_2$ \cite{Huang16}, and CaP$_3$ family \cite{Xu17}.
DC conductivities are calculated in hyperhoneycomb lattices with the TNLs \cite{Mullen15, Ezawa16}.
Recently, ZrSiS has been observed as a TNLSM experimentally \cite{Schoop16, Neupane16, Hu16, Wang16A, Singha17}.

Meanwhile, one can find another type of  spinless nodal lines protected by mirror or glide symmetry but not topology.
The nodal lines lie on the mirror/glide plane.
The spinless nodal-line semimetals have been reported in CaAgP and CaAgAs, which are noncentrosymmetric \cite{Yamakage16, Okamoto16}.
In general, the protection by the mirror symmetry can coexist with the topological protection.
Actually, the nodal lines in ZrSiS are protected also by the glide symmetry \cite{Schoop16, Neupane16}.

We can also realize the WSM phase in spinless systems
when either TR or I symmetry is absent \cite{Manes12, Fang12, Lu13, Zou16}.
%The spinless WSM can be regarded as the spinful Dirac semimetal whose nodes lie at general points, by including the spin degree of freedom. 
The spinless WSM phase has been realized experimentally in a photonic crystal \cite{Lu15, Chen15a}.
Some spinless WSM phases appear
between topologically trivial and nontrivial insulator phases characterized by some crystal symmetries \cite{Alexandradinata14, Kim16}.
In some models, 
the spinless WSM is expected to be driven from the nodal-line semimetal by a circularly polarized light \cite{Narayan16, Yan16, Chan16R, Taguchi16, Ezawa17, Yan17}. 

Yet the purely spinless TNLSMs and the spinless WSMs have not been discovered experimentally in three-dimensional electronic systems.
Additionally, the topological nodal-bands are suggested in non-electronic systems
which do not have spin-orbit interactions \cite{Lu13, Xiao15, Gao16, Rocklin16, Yang16, Li16c, Mook16, Mook17}.
Therefore, it is important to give a general framework of a phase transition between the spinless topological semimetal phases,
as studied in spinful systems \cite{Murakami07, Murakami08, Yang14}.
Moreover, it is recently shown that
band evolutions of spinful WSMs are determined by crystallographic symmetries, 
although the Weyl nodes may arise at generic points \cite{Murakami17}.
Extending this theory to the TNLSMs helps us to understand the band structures easily.

In this work, we study a generic topological phase transition between the spinless semimetal phases.
%Our purpose is to give a universal diagram for the topological semimetals in spinless systems.
To elucidate the phase transition, we classify TNLs into two types, type-A and type-B. 
The type-A and type-B TNLs are distinguished by their locations and shapes, which roughly corresponds to  whether or not the TNLs enclose a time-reversal invariant momentum. We show that depending on the type of the TNL, its topological nature and its evolution under symmetry-breaking
perturbations are quite varied. 
It is shown that the type-A TNLSM phase always becomes the spinless WSM phase when the TR symmetry is broken.
Furthermore, we show how other crystallographic symmetries
constrain positions of the TNLs in the type-A TNLSMs.
As a result, even if the I symmetry is broken, the system remains in a nontrivial topological semimetal phase by the crystal symmetries in many cases. 
We also demonstrate the phase transition between the TNLSM phase and the WSM phase by using a lattice model to confirm our theory.

This paper is organized as follows.
We classify the topological nodal line into the type-A and the type-B TNLSMs and 
show corresponding effective models in Sec.~\ref{effH}.
In Sec.~\ref{pt1}, we show general phase transitions in TNLSMs when the TR or I symmetry 
is broken. %phase to the spinless WSM phase.
We elucidate effects of other crystal symmetries on band structures and nodal lines of the topological semimetals in Sec.~\ref{stsp}.
In Sec.~\ref{ptcry}, we discuss phase transitions in TNLSMs with additional crystallographic symmetries, when the  the TR or I symmetry 
is broken.
Our results are summarized in Sec.~\ref{conclusion}.

\section{CLASSIFICATION OF TNLSMS INTO TWO TYPES}\label{effH}
In this section we classify TNLs into two types: type-A and type-B.
We consider systems with I and TR symmetries, whose 
operators are denoted by $P$ and $\Theta$, respectively. $\Theta$ is a complex conjugation operator $K$.
These symmetries give constraints $H(-\bm{k})=PH(\bm{k})P^{-1}=\Theta H(\bm{k})\Theta ^{-1}$, where $H(\bm{k})$ is the Hamiltonian.
Because of these symmetries, it is important to describe behaviors of the energy bands at a time-reversal invariant momentum (TRIM).

\begin{figure}[t]
\includegraphics[width=7.5cm]{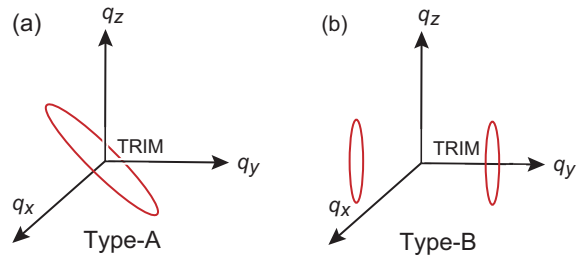}
\caption{\label{Type-AandB}
Schematic drawing of (a) type-A and (b) type-B TNLs. (a) When a type-A TNL encloses the TRIM,
the type-A TNL intersects with any of the TR-invariant planes including the TRIM $2(2N+1)$ times.
(b) A type-B TNL intersects with the TR-invariant plane $4N$ times.
}
\end{figure}

We classify TNLs based on their shapes around one of the TRIM.
Because of the TR symmetry, TNLs appear symmetrically with respect to TRIM.
When there are more than one TNL in the Brillouin zone, 
we consider each TNL separately.  
It may sometimes happen that a single TNL is not time-reversal invariant in itself, i.e. 
it is not symmetric with 
respect to the TRIM considered; 
in such cases, we consider instead a pair of TNLs which is symmetric with 
respect to the TRIM, as shown in Fig.~\ref{Type-AandB}(b). Obviously, this pairing of TNLs is independent of the 
choice of the TRIM. 
It may also happen that some TNLs may traverse across the Brillouin zone, like an ``open orbit'' 
of an electron under a magnetic field within semiclassical theory. Our theory also works in such cases.

To classify individual 
TNLs, we first define a TR-invariant plane, as a plane in $\bm{k}$ space containing the TRIM considered. 
This plane is invariant under the TR symmetry. 
Since the TNL are symmetric with respect to the TRIM,
the TNL always intersects with the TR-invariant plane $2(2N+1)$ or $4N$ times, where $N$ is a non-negative integer (Fig.~\ref{Type-AandB}).
If the TNL intersects with the TR-invariant plane $2(2N+1)$ times,
the TNL encloses the TRIM as shown in Fig.~\ref{Type-AandB}(a), and 
we call the TNL a type-A TNL.
On the other hand, if the number of the intersection points is $4N$,
we call the TNL a type-B TNL  as shown in Fig.~\ref{Type-AandB}(b).
If the TNLs are tangential to the TR-invariant plane, we slightly move the TR-invariant plane 
to eliminate the points of tangency, and count the number of intersections. 
This classification is independent of the choice of the TR-invariant plane for the fixed choice of the 
TRIM. Furthermore, it is also independent of the choice of the TRIM, which can be
directly shown by considering a TR-invariant plane containing more than one TRIM.

In the following, we construct a two-band effective Hamiltonian consisting of the conduction and the valence bands around the TRIM, 
in order to facilitate our understanding of the behaviors of the TNLs.
To construct an effective Hamiltonian we assume that each TNL is isolated, meaning that we
can take a vicinity of the TRIM which contain only one TNL.
First, we consider the case where
the parity eigenvalues of the conduction and the valence bands are different at the TRIM, and are inverted from 
the other TRIM. As we see later, this corresponds to the type-A TNL.
Then the I symmetry is given by $P=\pm \sigma _z$, 
where $\sigma _{i=x,y,z}$ denote Pauli matrices acting on the space spanned by the conduction and the valence bands.
Then, from the TR- and I symmetries, the effective Hamiltonian is
\begin{align}
H_{\mathrm{TNL}}(\bm{q})=a_y(\bm{q})\sigma _y+a_z(\bm{q})\sigma _z, \label{TNL1}
\end{align}
where $\bm{q}$ is a wavevector measured from the TRIM, $a_y(-\bm{q})=-a_y(\bm{q})$, and $a_z(-\bm{q})=a_z(\bm{q})$.
Therefore, the TNL is represented by $a_y(\bm{q})=0$ and $a_z(\bm{q})=0$. 
Here, we are considering the case 
where the parities of the bands at the TRIM are inverted from 
those at other TRIM. Therefore, the
coefficient $a_z(\bm{q})$ changes sign as we go away from the TRIM ($\bm{q}=0)$. Hence, the 
equation $a_z(\bm{q})=0$ defines 
a closed surface encircling the TRIM, and together with the other condition $a_y(\bm{q})=0$, it indeed defines a
TNL enclosing the TRIM, corresponding to the type-A TNL.
 In this case, a sign of a parameter $m$ 
defined by $m\equiv a_z(\bm{q}=0)$ describes whether the bands are inverted or not. Suppose we start from the TNLSM phase
and change this parameter $m$ across zero. As $m$ approaches zero, the nodal line shrinks. At $m=0$ the gap closes at the TRIM
 ($\bm{q}=0$), and then the gap opens. 

In addition, some of the type-B TNLs can also be described by Eq.~(\ref{TNL1}).
It happens when the sign
of $a_z(\bm{q})$ at $\bm{q}=0$ and that away from $\bm{q}=0$ are the same, whereas  $a_z(\bm{q})$ vanishes at some $\bm{q}$. This corresponds to the type-B TNL, by counting 
the number of intersections between the TNL and the TR-invariant plane.

Second,  when the parity eigenvalues of the conduction and the valence bands are identical at the TRIM, 
$P=\pm \sigma _0$ and the effective Hamiltonian is
\begin{align}
H_{\mathrm{TNL}}(\bm{q})=a_x(\bm{q})\sigma _x+a_z(\bm{q})\sigma _z, \label{TNL2}
\end{align}
where $a_x(-\bm{q})=a_x(\bm{q})$ and $a_z(-\bm{q})=a_z(\bm{q})$.
TNLs exist if $a_x(\bm{q})=0$ and $a_z(\bm{q})=0$. It is straightforward to see that 
the number of intersections between the TNL and the TR-invariant plane 
is $4N$ ($N$: integer), meaning that this TNL is of type B.
Unlike Eq.~(\ref{TNL1}), the gap closing  at the TRIM is prohibited by level repulsion. 
Meanwhile, as we explained later, the TNL can be annhilated without crossing the TRIM.

In some cases, there are more than one TNLs in the Brillouin zone. 
Ca (calcium) has four type-A TNLs and and Yb (ytterbium) without the spin-orbit coupling has six pairs of 
type-B TNLs \cite{Hirayama17}. 
Let $n_{\rm A}$ and $n_{\rm B}$ denote the number of type-A TNLs and that of type-B
TNLs, respectively. 
Then one can relate these numbers with 
the $\mathbb{Z}_2$ topological invariants $\nu_i$ $(i=0,1,2,3)$ introduced in Ref.~\onlinecite{Kim15}.
%Here, we discuss a relationship between the effective Hamiltonian and the topological invariants.
The topological invariants are defined as
\begin{align}
(-1)^{\nu _0}=\prod _{n_j=0,1}\prod _m^{occ.} \xi _{m}(\bm{\Gamma} _{n_1n_2n_3}), \label{ti0} \\
(-1)^{\nu _i}=\prod _{n_i=1,n_{j\neq i}=0,1}\prod _m^{occ.} \xi _{m}(\bm{\Gamma} _{n_1n_2n_3}), \label{ti}
\end{align}
where $\xi _m(\bm{\Gamma} _{n_1n_2n_3}) $ is a parity eigenvalue of the $m$-th occupied band 
at a TRIM $\bm{\Gamma} _{n_1n_2n_3} =(n_1\bm{G}_1+n_2\bm{G}_2+n_3\bm{G}_3)/2$, $n_i=0,1$. 
$\bm{G}_{i=1,2,3}$ are reciprocal vectors.
These topological invariants determine whether the number of intersections between the TNLs and a half of an arbitrary plane including four TRIM is even or odd \cite{Kim15}.
In particular, it directly follows from Ref.~\onlinecite{Kim15} that
\begin{align}
\nu_0\equiv n_{\rm A}\ ({\rm mod}\ 2).
\label{nu0nA}\end{align}
%We note that if the spin-orbit interaction is added, $(\nu _0; \nu_1 \nu_2 \nu_3)$ coincide with the Fu-Kane-Mele invariants \cite{Fu07}.
%When any of the topological invariants is equal to 1, the energy bands have the type-A TNLs
%because one of the TRIM has a different parity eigenvalue from the other TRIM.
%When one of the TRIM has a different parity eigenvalue from the rest of the TRIM, 
%the energy bands have the type-A TNL, which follows from Ref.~\onlinecite{Kim15}. Otherwise, 
%when the parity eigenvalue at the TRIM considered is the same with that at other TRIM, the TNL 
%s of type-B, from Ref.~\onlinecite{Kim15}. Thus, a type-B TNL is $\mathbb{Z}_2$ trivial while a type-A TNL is $\mathbb{Z}_2$ nontrivial.
%Since the change of the topological invariants represents the band inversion at a certain TRIM as discussed above,
%the energy bands necessarily have the type-A TNL in the nontrivial phase.
%We again emphasize that these $\mathbb{Z}_2$ topological numbers corresponds to the classification of the TNLs as a single set in the whole Brillouin zone, and it is not for an individual TNL. This makes a difference when 
%there are more than one TNLs in the Brillouin zone. 
For example, both in Ca and Yb (without the the spin-orbit coupling),  the $\mathbb{Z}_2$ topological numbers are trivial, i.e. $(\nu _0; \nu_1 \nu_2 \nu_3)=(0;000)$ \cite{Hirayama17}, and it agrees
with the number of TNLs, $(n_{\rm A},n_{\rm B})=(4,0)$ in Ca and $(n_{\rm A},n_{\rm B})=(0,6)$ in Yb.
%Namely, the trivial topological invariants does not necessarily indicate nonexistence of the type-A TNLs.

Next we consider an evolution of a TNL under continuous deformation of the system. 
A TNL may change its shape under the deformation, and sometimes the number of intersections with 
a TR-invariant plane may change.
We first note that as long as the TNL does not go across the TRIM, 
the number of intersections between a TNL and a TR-invariant plane can change 
only by an integer multiple of four, because the TNL remains symmetric with respect to the TRIM. 
Therefore, a type-B TNL can shrink and be annihilated without crossing the TRIM, because
$4N\equiv 0\ (\text{mod}\ 4)$. On the other hand, to annihilate
a type-A TNL, it should go across the TRIM, and thereby the gap closes at the TRIM. 
From the argument of the $\mathbb{Z}_2$ topological numbers, 
in order to annihilate a type-A TNL, the $Z_2$ topological number should change, and thus this gap closing 
should necessarily accompany an exchange of the parity eigenvalues at the TRIM
between the valence and the conduction bands. This agrees with the argument in Eq.~(\ref{TNL1}).
If the two bands forming the TNL have the same parity eigenvalues, 
the gap closing at the TRIM is not allowed because of the level repulsion.
Meanwhile, when the bands have opposite parity eigenvalues, there are no constraints for gap-closing points.
We remark that one can change the numbers of type-A TNLs and type-B TNLs
under continuous deformation of the system without changing the topological invariant $\nu_0$. 
For example, one can continuously deform from the TNLs in Ca to those in Yb via Lifshitz transitions, 
without changing $\nu_0$.

\section{PHASE TRANSITION INVOLVING TNLSMS}\label{pt1}
To elucidate phase transitions involving the TNLSM phase, 
we add symmetry-breaking perturbations to the system. 
We use the effective Hamiltonians for the TNLSMs described by Eqs.~(\ref{TNL1}) and (\ref{TNL2}).
We assume that the TNLs  are realized before breaking the symmetry, and let ${\bf \bm{\ell}}$ denote the TNL. 
On the TNLs, $a_y(\bm{q})=a_z(\bm{q})=0$ in Eq.~(\ref{TNL1}) or $a_x(\bm{q})=a_z(\bm{q})=0$ in Eq.~(\ref{TNL2}) holds.
In this section, we ignore crystallographic symmetries other than I symmetry.

\subsection{Type-A TNLSMs with TR breaking}
Firstly, we break the TR symmetry in type-A TNLSMs.
The allowed perturbation term is $a_x(\bm{q})\sigma _x$ which satisfies $a_x(-\bm{q})=-a_x(\bm{q})$ because of the I symmetry.
We can assume that the perturbation is so small that the coefficients of Eq.~(\ref{TNL1}) remain zero on ${\bf \bm{\ell}}$ after the TR breaking. 
Thus, the gap closes when $a_x(\bm{q})=0,^{\exists} \bm{q}\subset {\bf \bm{\ell}}$ in the presence of the small TR breaking term.
In fact, such wavevectors satisfying $a_x(\bm{q})=0$ always exist somewhere on ${\bf \bm{\ell}}$
because $a_x(\bm{q})$ is an odd function of $\bm{q}$ and the type-A TNL ${\bf \bm{\ell}}$ encloses the TRIM represented by $\bm{q}=0$. 
The emergent gapless points are Weyl nodes [Fig.~\ref{PT} (a)].
The Weyl nodes appear symmetrically with respect to the TRIM, 
and the minimal number of Weyl nodes is two.
The two Weyl nodes are related by the I symmetry, and thus have opposite monopole charges.
Hence, when the TR symmetry is broken, the system changes from the type-A TNLSM phase to the spinless WSM phase.

We also show another proof of the appearance of the spinless WSM phase by breaking the TR symmetry based on a topological description.
We assume that a type-A TNL encloses a TRIM $\bm{\Gamma}$, and that the energy bands are gapless only on the TNL.
We consider a TR-invariant plane $P_{\bm{\Gamma}}$ which includes $\bm{\Gamma}$.
The TR-invariant plane has $2(2N+1)$ intersection points $\pm \bm{k}_i (i=1, \cdots ,2N+1)$ with the type-A TNL.
We focus on pairs of the gapless points on $P_{\bm{\Gamma}}$, which are 
related by the I symmetry.
Because the closings of the gap at these gapless points 
are protected topologically by the TR- and I symmetries, 
the bands generally become gapped at these points when we weakly break the TR symmetry.
The perturbation terms obtained in each pair $\pm \bm{k}_i$ have opposite signs, because of the I symmetry.
%since the bottom of the conduction and the top of the valence bands have opposite parity eigenvalues.
Thus, the bands at each pair of wavevectors $\pm \bm{k}_i$ contribute by $+1$ or $-1$ to the Chern number defined on the plane 
$P_{\bm{\Gamma}}$ \cite{Haldane88}.
By summing over all the $(2N+1)$ pairs,
the Chern number on the plane  $P_{\bm{\Gamma}}$ is nonzero.
On the other hand, we introduce another plane $P_{{\bm{\Gamma}}\parallel }$ which is parallel to $P_{\bm{\Gamma}}$,
but does not intersect nodal lines [Fig.~\ref{PT} (b)].
By assumption, the Chern number defined on $P_{{\bm{\Gamma}}\parallel }$ is zero before introducing the perturbation. 
As long as the perturbation is small,
the band gap does not close on the plane $P_{{\bm{\Gamma}}\parallel }$,
and the Chern number remains zero on $P_{{\bm{\Gamma}}\parallel }$ after the TR symmetry is broken.
Therefore, the Chern numbers are different between $P_{\bm{\Gamma}}$ and $P_{{\bm{\Gamma}}\parallel }$,
and it means that between $P_{\bm{\Gamma}}$ and $P_{{\bm{\Gamma}}\parallel }$ the energy bands should have gapless points, i.e. Weyl nodes.
As a consequence, the WSM phase necessarily emerges from the type-A TNLSM phase by breaking the TR symmetry.

\subsection{Type-A TNLSMs with I breaking}
Secondly, we introduce a term which weakly breaks the I symmetry but preserves the TR symmetry in type-A TNLSMs.
The allowed term is described by $a_x(\bm{q})$ which satisfies $a_x(-\bm{q})=a_x(\bm{q})$.
Then, $a_x(\bm{q})$ can be nonzero on the whole loop ${\bf \bm{\ell}}$ since $a_x(\bm{q})$ is an even function of $\bm{q}$.
Therefore, the energy bands can become gapped.
It is natural from the viewpoint of topology; because the perturbation terms obtained in each pair $\pm \bm{k}_i$ have the same signs,
the Chern number on the plane $P_{\Gamma}$ is zero, implying that 
there appear no gapless points in general. 

\subsection{Type-B TNLSMs with TR or I breaking}
Next, we study a phase transtion of the type-B TNLSM phase by breaking the TR- or I symmetry. 
The additional perturbation term for Eq.~(\ref{TNL1}) and (\ref{TNL2}) for breaking either of the TR- or I symmetries 
is $a_x(\bm{q})\sigma _x$ and $a_y(\bm{q})\sigma _y$, respectively. 
Now the perturbation is generally nonzero everywhere on ${\bf \bm{\ell}}$, whichever symmetry is broken.
Even if the perturbation term is an odd function of $\bm{q}$,
it can be nonzero on ${\bf \bm{\ell}}$ because the type-B TNLs do not enclose the TRIM,
unlike the type-A TNLs.
Therefore, in general, by breaking the TR- or I symmetry, a gap opens, and the WSM phase does not appear from the type-B TNLSM phase.

\subsection{Phase transition in a lattice model}
In this subsection, we see a phase transition from the type-A TNLSM phase to the spinless WSM phase by using a lattice model, 
and we see agreement with the discussion in Sec.~\ref{pt1}A.
%Actually, we show that the phase transition occurs by breaking TR symmetry even without considering additional crystal symmetries.
We use a model on a diamond lattice given by
\begin{align}
H=\sum _{<ij>}t_{ij}c^{\dagger}_ic_j +\sum _{\ll ij\gg }t'_{ij}{c^{\dagger}_i}c_j.
\end{align}
The first term represents nearest-neighbor hoppings between the sublattices A and B.
Here, we denote the three translation vectors by $\bm{t}_1=\frac{a}{2}(0,1,1),\bm{t}_2=\frac{a}{2}(1,0,1)$, and $\bm{t}_3=\frac{a}{2}(1,1,0)$,
where $a$ is a lattice constant.
Then, the four nearest-neighbor bonds are $\bm{\tau}=\frac{a}{4}(1,1,1)$, and $\bm{\delta}_{i=1,2,3}=\bm{\tau}-\bm{t}_{i=1,2,3}$.
%$\bm{\delta} _1=\frac{a}{4}(1,-1,-1),\bm{\delta} _2=\frac{a}{4}(-1,1,-1),\bm{\delta} _3=\frac{a}{4}(-1,-1,1),\bm{\tau }=\frac{a}{4}(1,1,1)$,
We express the hoppings in the direction of $\bm{\delta}$ as subscripts.
For example, the hoppings in the direction of $\bm{\tau}$ and $\bm{\delta} _{i=1,2,3}$ are written by $t_{\bm{\tau}}$ and $t_{\bm{\delta} _{i=1,2,3}}$, respectively.
%For simplicity, we assume that the hoppings are independent of the direction.
The second term represents the next nearest-neighbor hoppings.
The twelve next nearest-neighbor bonds are represented by $\pm \bm{t}_{i=1,2,3}$, and $\pm \bm{u}_{i=1,2,3}$,
where $\bm{u}_1=\bm{t}_3-\bm{t}_2, \bm{u}_2=\bm{t}_3-\bm{t}_1$, and $\bm{u}_3=\bm{t}_1-\bm{t}_2$.
In addition, we denote the next nearest-neighbor hoppings between the same sublattices A(B) by $t'^{A(B)}_{\bm{\delta}}$.
When the system is I-symmetric, $t_{\bm{\tau}}$ and $t_{\bm{\delta} _{i=1,2,3}}$ are real, and $t'^A_{\bm{\delta}}=(t'^B_{\bm{\delta}})^{\ast}$.
The Hamiltonian in the momentum space is
\begin{widetext}
\begin{align}
H(\bm{k})=
\Bigl[2\sum_{\bm{d}} \mathrm{Re}[t'^A_{\bm{d}}]\cos \bm{k}\cdot \bm{d}\Bigr] \sigma _0+
\Bigl[ t_{\bm{\tau}}+\sum _it_{\bm{\delta}_i}\cos \bm{k}\cdot \bm{t}_i\Bigr] \sigma _x+
\Bigl[\sum _i t_{\bm{\delta}_i}\sin \bm{k}\cdot \bm{t}_i\Bigr] \sigma _y+
\Bigl[ 2\sum_{\bm{d}} \mathrm{Im}[t'^A_{\bm{d}}]\sin \bm{k}\cdot \bm{d}\bigr] \sigma _z,
%a_0(\bm{k})\sigma _0+a_x(\bm{k})\sigma _x+a_y(\bm{k})\sigma _y+a_z(\bm{k})\sigma _z,
\end{align}
\end{widetext}
where $\bm{d}$ in the sum runs over $\bm{t}_{i=1,2,3}$ and $\bm{u}_{i=1,2,3}$.
The Pauli matrices $\sigma _{i=0,x,y,z}$ act on the sublattice degree of freedom.
In this model, the parity operator is represented by $P=\sigma _x$.
Then, the parity eigenvalues $\xi$ of the occupied bands at the TRIM $\bm{\Gamma}_{n_1n_2n_3}$ are given by
\begin{align}
\xi (\bm{\Gamma}_{n_1n_2n_3})=-\mathrm{sgn}\Bigl[ t_{\bm{\tau}}+\sum _it_{\bm{\delta}_i}(-1)^{n_i}\Bigr] .
\end{align}
The topological invariant $\nu _0$ is obtained from $(-1)^{\nu _0}=\prod _{n_j=0,1}\xi (\bm{\Gamma}_{n_1n_2n_3})$.

\begin{figure}[h]
\includegraphics[width=7.5cm]{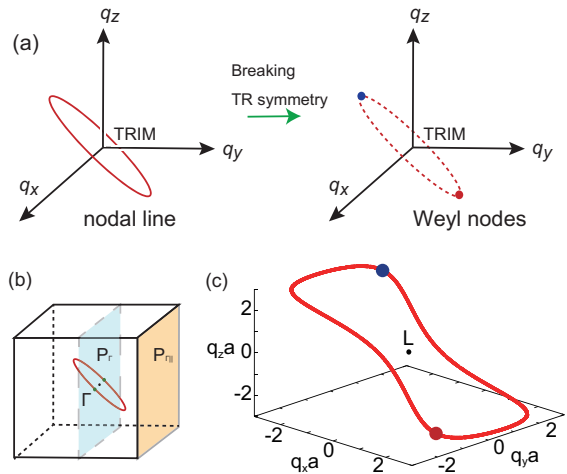}
\caption{\label{PT}
(a) Schematic drawing of the band evolution.
The red solid line is the TNL around the TRIM.
The blue and the red points are the spinless Weyl nodes left on the loop ${\bf \bm{\ell}}$ (red dashed line).
The difference between colors of the Weyl nodes corresponds to opposite monopole charges.
(b) Schematic drawing of the type-A TNL and the time-reversal invariant planes. 
The red line is the type-A TNL.
The blue and orange regions represent $P_{\bm{\Gamma}}$ and $P_{{\bm{\Gamma}}\parallel }$, respectively. 
The green dots are the intersection points of the type-A TNL and $P_{\bm{\Gamma}}$.
(c) Change of the band structure from the type-A TNL to the spinless Weyl nodes in a lattice model.
The axes represent wavevectors measured from the point $L$.
The red loop is the type-A TNL around the $L$ point
when $t_{{\bm{\tau}}}/t_{\bm{{\delta} _1}}=1.4, t_{\bm{\delta }_2}/t_{\bm{\delta} _1}=1.1$, $t_{\bm{\delta }_3}/t_{\bm{\delta} _1}=0.9$, and $t/t_{\bm{\delta} _1}=0.1$.
The blue and red dots are the Weyl nodes which appear from the type-A TNL for the finite TR breaking $\phi =0.1$ .
}
\end{figure}

When $t'^{A(B)}_{\bm{\delta}}$ are real i.e. $\mathrm{Im}[t'^A_{\bm{\delta}}]= 0$, the model has TR symmetry, which case has
been studied in Ref.~\onlinecite{Takahashi13}.
In Ref.~\onlinecite{Takahashi13}, it is shown that the energy bands can have a type-A TNL around the TRIM $\bm{\Gamma}_{111}=L=\frac{\pi}{a}(1,1,1)$.
The type-A TNL exists when the parity eigenvalue $\xi (\bm{\Gamma}_{111})$ is opposite from those at the other TRIM.
To realize it, we set $t_{{\bm{\tau}}}/t_{\bm{{\delta} _1}}=1.4, t_{\bm{\delta }_2}/t_{\bm{\delta} _1}=1.1$, and $t_{\bm{\delta }_3}/t_{\bm{\delta} _1}=0.9$.
We also assume that all the second nearest neighbor hopping are identical, having the values $t'^A_{\bm{\delta}}/t_{\bm{\delta} _1}=0.1$.
Since all the nearest-neighbor hoppings are different and real, the model has only TR- and I symmetries.
Then, the type-A TNL appears around the $L$ point as seen in Fig.~\ref{PT} (c), which 
is as expected from our argument in Sec.~\ref{pt1}A.

Next we break the TR symmetry by adding finite imaginary parts of $t'^A_{\bm{\delta}}$.
For example, this TR breaking can be included as Peierls phases from magnetization.
We put $t'^A_{\bm{d}}=te^{i\phi}$ for  the next nearest-neighbor hoppings  represented by $\bm{d}=\bm{t}_{i=1,2,3}$ and $\bm{u}_{i=1,2,3}$, where $t$ and $\phi$ are real constants. 
Then, $t'^A_{\bm{d}}=te^{-i\phi}$ when $\bm{d} =-\bm{t}_{i=1,2,3}$ and $\bm{d}=-\bm{u}_{i=1,2,3}$.
%Here, $t$ and $\phi$ are an absolute value and an argument of $t'^A_{\bm{\delta}}$.
In order to break the TR symmetry, we put $t/t_{\bm{\delta} _1}=0.1$ and $\phi =0.1$ for instance.
Consequently, we find that a topological phase transition occurs from the type-A TNLSM phase to the spinless WSM phase.
Figure \ref{PT} (c) shows the two Weyl nodes which emerge from the type-A TNL.

Instead of the TR-breaking, we can break the I symmetry by adding an on-site staggered potential given by
$H_{IB}=M_s\sum _i\lambda _ic_i^{\dagger}c_i$,
where $\lambda _i$ takes values $+1$ for the A sublattices and $-1$ for the B sublattices.
Then, we can directly see that the type-A TNL becomes gapped.

\section{CRYSTAL SYMMETRIES AND BAND STRUCTURES OF TYPE-A TNLSMs}\label{stsp}
In Sec.~\ref{pt1}, we have discussed the TNLSMs by considering only TR- and I symmetries.
In this section, we also take account into twofold rotational $(C_2)$ and mirror $(M)$ symmetries, because $C_2M $ is equal to the space inversion. 
Particularly, we show how the two symmetries, $C_2$ and $M$, constrain band structures having type-A TNLs.
%As a result, the constraint gives a simple way to understand band structures of the TNLSMs in many crystals.

\subsection{Band structures of the type-A TNLSMs}
Now, we classify the type-A TNLs into two cases according to whether or not the TRIM which the TNLs enclose is invariant under $C_2$ and $M$ symmetries. Because $P=C_2M$, in I-symmetric systems, a little group of the TRIM often contains $C_2$ and $M$ symmetries in pairs. 
If the TRIM is not invariant under the two symmetries,
we call this case (I). When the TRIM considered is invariant under $C_2$ and $M$, we call the case (II). 
Here, twofold screw symmetries and glide symmetries can be treated similarly to 
$C_2$ and $M$ symmetries, respectively, and the systems with these symmetries can be included in the case (II), except for some special cases at the 
Brillouin zone boundary for nonsymmorphic space groups (see the Appendix \ref{classsym}).
Actually, the case (II) 
is more  important for application to real materials
because 89 space groups of all the 92 space groups with I symmetry have the two symmetries \cite{Bradley10}.

In fact, 
the band structures and the phase transition of the type-A TNLSMs for (I) have already 
been studied in Sec.~\ref{effH} and \ref{pt1}, 
because there is no additional symmetry which further constrains the phase transition. 
For example, CaP$_3$ \cite{Xu17} is included in the case (I).

\begin{figure}[t]
\includegraphics[width=7.5cm]{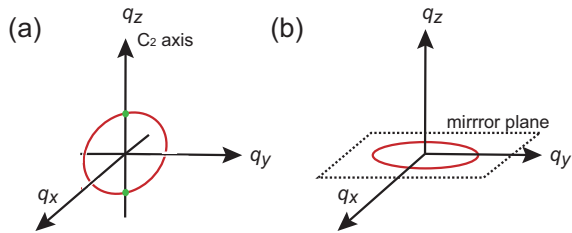}
\caption{\label{linesym}
Schematic positions of the nodal lines in the type-A TNL around the TRIM.
(a) The type-A TNL for the case (i) has the intersection points with the $C_2$ invariant axis,
which are indicated by the green dots.
(b) For the case (ii), the type-A TNL appears on the mirror-invariant plane.
}
\end{figure}

In the case (II), the two symmetries $C_2$ and $M$ give some constraints to the effective Hamiltonian described by Eq.~(\ref{TNL1}). 
Since the parity eigenvalues are different for the conduction and the valence bands of the type-A TNLs,
either of the $C_2$ or the $M$ symmetry has different eigenvalues for the conduction and the valence bands.
%Thus, it leads to case (II) because either of the $C_2$ or the $M$ symmetry gives the different eigenvalues to the conduction and the valence bands. 
%Therefore, in such cases, the type-A TNLs necessarily cross on the $C_2$-invariant axes or lies on the mirror plane in this case.
In spinless systems, eigenvalues of the $C_2$ and the $M$ symmetries take values $\pm 1$.
Then, because $P=MC_2$, there are two cases for combinations of eigenvalues $C_2$ and $M$ at the TRIM;
(II)-(i) eigenvalues of $M$ are the same and those of $C_2$ are different, and 
(II)-(ii) eigenvalues of $M$ are different and those of $C_2$ are the same.
They correspond to two different matrix representations: (II)-(i) $M=\pm \sigma _0$ and $C_2=\pm \sigma _z$, and (II)-(ii) $M=\pm \sigma _z$ and $C_2=\pm \sigma _0$.
We can calculate the band structures for these cases, and the details are shown in the Appendix \ref{classsym}.
The resulting positions of the nodal lines are shown in 
Fig.~\ref{linesym}, where we set the twofold rotational axis and the mirror plane to be the $z$ axis and $xy$ plane, respectively.
For (II)-(i), as seen in Fig.~\ref{linesym} (a), the type-A TNL encircles the TRIM and intersects the $C_2$-invariant axis.
The TNL is symmetric with respect to the mirror plane $q_z=0$.
For (II)-(ii), the type-A TNL appears on the mirror-invariant plane as shown in Fig.~\ref{linesym} (b).

\subsection{Applications of  the theory of the band structures to the candidate materials
of TNLSMs}
The results in the previous subsection can be easily generalized to little groups with many different pairs of $C_2$ and $M$ symmetries.
We apply the theory to several candidates of the type-A TNLSMs.
First, we consider fcc Ca, whose space group is No.~225 \cite{Hirayama17}.
Ca have four type-A TNLs near each of the four TRIM $L$.
The little group at the $L$ points is $D_{3d}$ which contains three $C_2$-rotational operations.
The two bands forming the TNLs belong to $A_{1g}$ and $A_{2u}$ states at the point $L$,
and they have different $C_2$ eigenvalues.
Therefore, the TNLs in Ca intersect the $L$-$W$ lines, which are the $C_2$ invariant axes 
but do not lie on mirror planes, in accordance with our theory.

Next, we apply this theory to Cu$_3$ZnN \cite{Kim15}.
The space group of Cu$_3$ZnN is No.~221.
The energy bands have type-A TNLs around the three TRIM $X$, whose little groups are $D_{4h}$.
The type-A TNLs are formed by $A_{2u}$ and $A_{1g}$ states at the $X$ points.
The $D_{4h}$ group contains a fourfold-rotational ($C_{4}$) operation, four $C_2$ operations whose rotational axes are normal to the principal axis,
and the corresponding five mirror operations.
Then, the $A_{2u}$ and $A_{1g}$ states have opposite eigenvalues of the four $C_2$ symmetries.
Thus, the TNLs cross the $X$-$M$ lines and $X$-$R$ lines, which are the $C_2$-invariant axes.
Meanwhile, the two states have the same eigenvalues of the $C_4$ symmetry.
Hence, $C_4=\sigma _0$ leads to different eigenvalues of the mirror symmetry $M=P(C_4)^2=\sigma _z$ .
Therefore, the TNLs also appear on the mirror-invariant plane normal to the $C_4$-invariant axis.
As a result, the type-A TNLs in Cu$_3$ZnN not only cross the $C_2$-invariant axes but also exist on the mirror plane.

Last, we remark that the type-A TNLs are predicted to appear on the mirror planes in many candidates
such as Cu$_3$(Pd, Zn)N \cite{Kim15, Yu15}, Ca$_3$P$_2$ \cite{Xie15, Chan16}, LaN \cite{Zeng15}, and compressed black phosphorus \cite{Zhao16}. They belong to (II)-(ii) and 
the existence of 
TNLs is understood from the difference in eigenvalues of $M$ between the conduction and valence bands.

\section{PHASE TRANSITIONS OF TYPE-A TNLSMS AND CRYSTAL SYMMETRIES}\label{ptcry}
In this section, we show that for type-A TNLSMs in the case (II), the presence of
$C_2$ and $M$ symmetries changes phase transitions when we break TR- or I symmetry.
%In the case (II)-(i), the $C_2$ symmetry allows a WSM phase even if I symmetry is broken.
%On the other hand, in the case (II)-(ii),
%nodal lines can survive without TR- and I symmetries.

\subsection{Type-A TNLs protected by crystal symmetries with TR breaking}
\begin{figure}[t]
\includegraphics[width=7cm]{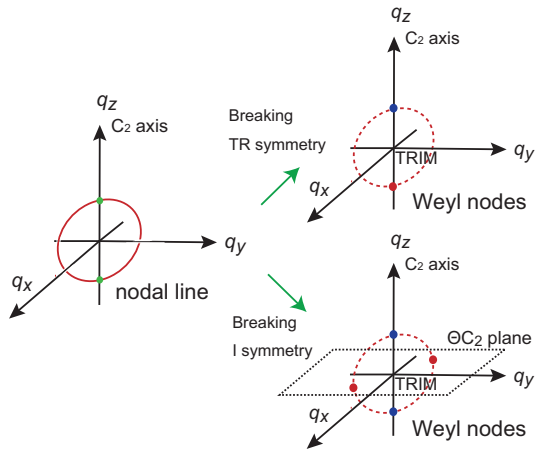}
\caption{\label{c2wsm}
Realization of the spinless Weyl nodes by breaking the TR or the I symmetry in the case (II).
The blue and the red points represent the Weyl nodes.
If the TR symmetry is broken, the Weyl nodes appear on the $C_2$ invariant axis.
When the I symmetry is broken, we obtain a pair of Weyl nodes on the $C_2$-invariant axis and
another pair on the $\Theta C_2$-invariant plane.
}
\end{figure}

Here, we break the TR symmetry in the type-A TNL. When energy bands cross on high-symmetry lines or planes, and have different eigenvalues of crystal symmetries,
the band crossing is protected by the symmetries. Therefore, such degeneracy 
remains on high-symmetry lines or planes, even when the TR symmetry is broken.
%From Sec.~\ref{stsp}, a type-A TNL in the case (II) is formed by two bands which have opposite $C_2$ or $M$ eigenvalues at the TRIM.
%As a result, the type-A TNL intersects with the $C_2$ axis, or appear on the mirror plane.
%Therefore, the energy bands maintain gapless nodes on the $C_2$ axis or the mirror plane if the system breaks the TR symmetry.
Therefore, in the case (II)-(i), where the type-A TNL crosses the $C_2$-invariant axis,
the TR breaking creates Weyl nodes on the $C_2$-invariant axis as shown in Fig.~\ref{c2wsm}.
In the effective model, the protection originates from the fact that the perturbation $a_x(0,0,q_z)$ always vanishes on the $C_2$ axis. 

Next, in the case (II)-(ii), where the type-A TNL is always on the mirror-invariant plane, 
the nodal line remains on the mirror plane even without the TR symmetry. 
In the effective $2\times 2$ model, it is seen from the fact that the perturbation 
$a_x(q_x,q_y,0)$ vanishes on the mirror plane.
In particular, one needs to break the $M$ symmetry in order to realize the WSM phase.

\subsection{Type-A TNLs protected by crystal symmetries with I breaking}
We study effects of the I breaking for the case (II) in this subsection.
%The type-A TNLs for the case (II) are protected by crystal symmetries as well as the topology.
In the case (II), where the system has $C_2$ and $M$ symmetries, 
violation of the I symmetry is equivalent to breaking either $C_2$ or $M$ symmetry because $P=MC_2$.
Therefore, the topological semimetal phases may survive
in a different way between (II)-(i) and (II)-(ii).

First, we consider the case (II)-(i).
When the type-A TNL intersects $C_2$-invariant axes,
the system becomes a spinless WSM phase by breaking the I symmetry while retaining the $C_2$ symmetries.
Then, we obtain Weyl nodes not only on the $C_2$ invariant axes but also on the  $\Theta C_2$-invariant plane ($q_z=0)$,  because of the symmetry protection.
In the effective model, 
$\Theta C_2=K\sigma _z$ symmetry leads to  $a_x(q_x,q_y,0)=0$, meaning that the perturbation is absent on this $\Theta C_2$-invariant  plane.
Here, within each pair of nodes related by the TR symmetry, monopole charges are
the same.
The four nodes correspond to the minimal number of Weyl nodes in TR-invariant WSMs 
\cite{Murakami07}. 
In fact, the appearance of four Weyl nodes can be understood by expanding the I-breaking perturbation term proportional to $\sigma _x$.
The term expanded near the TRIM to the lowest order is
$a_x(\bm{q})=(\alpha q_x+\beta q_y)q_z$.
%\begin{align}
%a_x(\bm{q})=(\alpha q_x+\beta q_y)q_z.
%H(\bm{q})&=(\alpha q_x+\beta q_y)q_z\sigma _x+(v_xq_x+v_yq_y)\sigma _y \nonumber \\
%&+(m+Aq_x^2+Bq_y^2+Cq_z^2+Dq_xq_y)\sigma _z.
%\end{align}
Therefore, we can see that the Weyl nodes appear when either $q_z=0$ or $q_x=q_y=0$ 
is satisfied, giving the four Weyl nodes.

On the other hand, for the case (II)-(ii) of the type-A TNLs on the mirror plane.
if we leave the $M$ symmetry and break the $C_2$ symmetry,
the nodal line survives because of the $M$ symmetry.
The mirror symmetry protects the nodal lines regardless of existence of the I symmetry. 
Hence, even if the I symmetry is broken, 
the nodal line remains unless the mirror symmetry is broken.

\section{CONCLUSION AND DISCUSSION}\label{conclusion}
In the present paper, we study phase transitions and band evolutions of topological nodal-line semimetals.
We classified topological nodal-line semimetals into type-A and type-B in order to describe general phase transitions by breaking time-reversal or inversion symmetry.
This classification is based on the geometrical positions of the nodal lines, and 
we give effective Hamiltonians for each case for analysis of symmetry breaking. 
The results show that the topological nodal lines enclosing a TRIM 
(type-A topological nodal lines)
always become Weyl nodes when the time-reversal symmetry is broken.
However, breaking of inversion symmetry opens a band gap in the type-A topological nodal-line semimetals, and 
it is confirmed by our calculation on the lattice model.
On the other hand, it is shown that the type-B topological nodal lines, which do not enclose a TRIM, 
become gapped by breaking time-reversal symmetry.
The two types are distinguishable from the shapes of the topological nodal lines.

We also showed how band structures of type-A topological nodal lines are determined by the little group at the TRIM.
When the topological nodal line encircles the TRIM, which is invariant under $C_2$ and $M$ symmetries of the system,
they cross the $C_2$-invariant axis and/or appear on the mirror-invariant plane,
and consequently are protected by the symmetries. Therefore, the nodal lines or points
survive in some cases, even when the time-reversal or the inversion symmetries is broken.
The revealed properties are also useful to search spinless topological semimetals in many materials
because many space groups with I symmetry have various $C_2$ and $M$ symmetries.
As a result, the spinless WSMs can be predicted in many candidates of the topological nodal-line semimetals protected by the $C_2$ symmetries
when we break the I symmetry.

Our study tells us how to realize a spinless WSM phase.
In electronic systems, the spinless WSM phase appear from the type-A topological nodal-line semimetal phases 
not only by a circularly polarized light \cite{Narayan16, Yan16, Chan16R, Taguchi16, Ezawa17, Yan17}
but also by magnetic ordering, an external electric field, structural transition, and so on.
Moreover, our theory can be applied to spinless fermions in cold atoms and bosonic bands.
The experiments have potential in bosonic metamaterials of photons and phonons where lattice structure and its symmetry are flexibly controllable.

The TNLs may cross each other, and our classification into type-A and type-B still works for the TNLs with mutual crossings. Meanwhile, in the presence of crossings, 
the effective models become different from those discussed in our paper, which is beyond the scope of this paper. In this context, classification of 
possible patterns of their crossings  and their evolution under 
the spin-orbit coupling was studied recently in Ref.~\onlinecite{Kobayashi17}. 
In Ref.~\onlinecite{Kobayashi17} only the TNLs based on the mirror symmetry are discussed, 
and the main focus is on the crossings of the TNLs. Meanwhile our paper includes both
the TNLs from the mirror symmetry and those from the $\pi$ Berry phase, and therefore
the target of our research is different from that of Ref.~\onlinecite{Kobayashi17}. 

\begin{acknowledgments}
This work was supported by JSPS KAKENHI Grant Numbers 16J08552 and 26287062
and by MEXT Elements Strategy Initiative to Form Core Research Center (TIES).
\end{acknowledgments}

\appendix
\section{Classification of the type-A TNLs and their band structures}\label{classsym}
We have classified type-A TNLs into the two cases (I) and (II) in Sec.~\ref{stsp}.
The classification provides information on band evolutions and phase transitions involving type-A TNLs.
In this appendix, we explain band structures of type-A TNLs in the two cases
(I) and (II).
%To see the band structures, we consider only type-A TNLs and a TRIM enclosed by the TNLs.
The type-A TNLs are formed by two bands with opposite parity eigenvalues at the TRIM.
%Namely, we assume that the type-A TNLs exist around the TRIM, and they can shrink to the TRIM continuously.
and they can be descrived by the two-band effective Hamiltonian by Eq.~(\ref{TNL1}).
Here, to describe the TNLs by the two-band effective Hamiltonian,
we assume that the TNLs are formed only by two nondegenerate states.
In fact, a similar two-band effective Hamiltonian has been used in 
 spinful WSMs in order to describe band evolutions in Ref.~\onlinecite{Murakami17},
and therefore,  
here we can extend the analysis in Ref.~\onlinecite{Murakami17}
to some of the spinless TNLSMs as well.
In Ref.~\onlinecite{Murakami17},
it is shown that when two bands touch each other on high-symmetry lines or planes,
emergent gapless nodes evolve along the lines and the planes where the two bands have the different eigenvalues.
By using these results, we can understand band structures of type-A TNLs.

For example, in Sec.~\ref{stsp} we introduced two cases  (II)-(i) and (ii) for type-A TNLs, 
formed by two bands with $C_2$ and $M$ symmetry.  These two cases 
are classified according to the $C_2$ and $M$ eigenvalues at the TRIM, and 
the effective Hamiltonian for the two cases are constrained by these two symmetries.
For simplicity, we set the $C_2$ axis and the $M$ plane to be the $z$ axis and $xy$ plane, respectively.
From the constraints, we obtain $\sigma _z H_{\mathrm{TNL}}(-q_x,-q_y,q_z)\sigma _z=H_{\mathrm{TNL}}(q_x,q_y,q_z)$ in the case (II)-(i).
Meanwhile, in the case (II)-(ii), we obtain $\sigma _z H_{\mathrm{TNL}}(q_x,q_y,-q_z)\sigma _z=H_{\mathrm{TNL}}(q_x,q_y,q_z)$.
Although eigenvalues of the $C_2$ and $M$ symmetries are different in spinless and spinful systems,
the expressions for these constraints are the same both in spinless and in spinful cases \cite{Murakami17}.
As a result, the type-A TNL of the case (II)-(i) crosses the $C_2$ axis while the type-A TNL of the case (II)-(ii) appears on the $M$ plane, both in spinless and in spinful systems.
In some cases, several type-A TNLs can enclose the same TRIM if a little group at the TRIM contains some $C_2$ and $M$ symmetries.

There are various options for the little group at the TRIM, which affects the position of the 
type-A TNLs. 
(I) refers to the case where the TRIM is neither $C_2$- nor $M$-symmetric.
Therefore, the little group is $C_i$ or $C_{3i}$.
In this case (I), the type-A TNL does not necessarily cross high-symmetry lines.
For example, if the two bands have the same $C_3$ eigenvalues at the TRIM whose little group is $C_{3i}$,
the type-A TNL lies at a general position.

(II) refers to the case where the TRIM is $C_2$- and $M$-symmetric.
In this case, as we have shown in Sec.~\ref{stsp}, 
the type-A TNLs necessarily cross the high-symmetry lines or appear on the mirror-invariant planes, thanks to symmetry protection.
The little groups can also have rotational symmetries besides the $C_2$ symmetry.
When the conduction and the valence bands belong to different subspaces of the $C_n$-rotational symmetries,
the type-A TNLs can intersect the high-symmetry lines.
On the other hand, when the two bands belong to different subspaces of the mirror symmetry,
the type-A TNLs are on the mirror-invariant plane.
In particular, if the eigenvalues of the $C_4$ or $C_6$ symmetry are the same,
the type-A TNL exists on the mirror-invariant plane perpendicular to the $C_4$- or $C_6$-invariant axes
because $(C_4)^2=C_2$ and $(C_6)^3=C_2$.

Here we comment on TNLs in systems with a nonsymmorphic space group having twofold screw $(S_2)$ symmetries or glide $(G)$ symmetries.
Inside the Brillouin zone, the TNLs are similar to those with a symmorphic space group,
because there is no extra degeneracy due to nonsymmorphic symmetry together with TR symmetry.
%since the eigenvalues of the $S_2$ and $G$ operations are $\pm 1$ at the TRIM.
Meanwhile, $S_2$ and $G$ symmetries may give rise to extra degeneracy on the Brillouin zone boundary by TR symmetry,
if the square of the symmetry operations becomes $-1$.
For example, by combination of several $G$ symmetries and the TR symmetry,
spinless nodal lines can contain fourfold-degenerate points at the 
TRIM on the surface of the Brillouin zone \cite{Takahashi17}.
As another example, a nodal surface appears on the $\Theta S_2$-invariant plane on the surface of the Brillouin zone \cite{Liang16}.
Such cases are beyond the scope of this paper because the band crossing cannot be described by the two-band effective Hamiltonian.
%Otherwise, since the nonsymmorphic symmetries can be represented by $\pm i\sigma _0$ or $\pm i\sigma _z$,
%we can treat the nonsymmorphic symmetries like $C_2$ and $M$ symmetries.

\bibliographystyle{apsrev4-1}
\bibliography{spinlesswsm}% Produces the bibliography via BibTeX.

\end{document}